\documentclass[proceedings]{JHEP37}
\usepackage{amssymb}
\usepackage{eurosym}
\usepackage{amsfonts}
\usepackage{amsmath,cite}
\usepackage{graphicx}
\usepackage{multirow}
\usepackage{epsfig}
\usepackage[title]{appendix}
\DeclareMathOperator\arccosh{arccosh}

\newbox\mybox

\newcommand\fverb{\setbox\mybox=\hbox\bgroup\verb}
\newcommand\fverbdo{\egroup\medskip\noindent\fbox{\unhbox\mybox}\ }
\newcommand\fverbit{\egroup\item[\fbox{\unhbox\mybox}]}
\conference{Moduli spaces for PT-regularized solitons}
\abstract{We construct and analyse the moduli space (collective coordinates) for a classical field theory in 1 + 1 dimensions that possesses complex stable multi-soliton solutions with real energies when PT-regularized. For the integrable Bullough-Dodd model we show, by comparing with the exact solutions, that a one-dimensional moduli space captures well the main feature of the centre of mass motion of the one and two-soliton solutions. We demonstrate that even the time-delay and spatial displacements occurring for the one-soliton constituents in a multi-soliton scattering process can be extracted from a moduli space analysis. We propose a two dimensional moduli space to describe the newly found triple bouncing scattering amongst the constituents of a dark two double peakon scattering.}

\title{Moduli spaces for PT-regularized solitons}
\author{Francisco Correa$^\circ$, Andreas Fring$^\bullet$ and Takanobu Taira$%
	^\ast$ \\
	$\circ$ Instituto de Ciencias F{\'{\i}}sicas y Matem{\'{a}}ticas,
	Universidad Austral de Chile, \\
	$\,\,$ Casilla 567, Valdivia, Chile\\
		$\bullet$ Department of Mathematics, City, University of London,\\
	$\,\,$ Northampton Square, London EC1V 0HB, UK \\
	$\ast$ Research Fellow of Japan Society for the Promotion of Science, Institute of Industrial \\ $\,\,$ Science, The University of Tokyo
	5-1-5 Kashiwanoha, Kashiwa 277-8574, Japan\\
	E-mail: francisco.correa@uach.cl, a.fring@city.ac.uk,
	taira904@iis.u-tokyo.ac.jp}

\input{tcilatex}
\begin{document}
	
\section{Introduction}
Moduli space constructions, also referred  to as collective coordinate methods, have been especially successful in the description of the manifolds of possible vacua in quantum field theories, e.g. \cite{seiberg1988observations,hanany1995quantum,argyres1996moduli} and in the characterisation of backgrounds in string theory, e.g. \cite{seiberg1994exact}. In the context of quantum field theories these methods have also turned out to provide extremely good insights into the dynamical properties of solitons \cite{takyi2016coll,manton2021kink}. In this application the advantage of the approach is that it reduces the complicated field equations to simpler classical equations of motion describing point particles. Many of the key features of the wave dynamics are well captured in the collective coordinate description by identifying the motion of the point particles with the centre of mass motion or distance of the waves. Some well-known prototype wave-equations such as variants of the Korteweg-de Vries equations reduce to well-known multi-particle systems such as the Calogero models \cite{Assis:2009gt}.   

Remarkably this approach does not only work well for integrable systems  \cite{caputo1991kink,manton2021kink}, when one can strictly speak of multi-solitons, but as well as for non-integrable systems  \cite{kivshar1989dynami,sugiyama1979kink,belova1988quasi,caputo1991kink,quintero2000ano,GoodHab2005,goodman2007chaotic,romanc2010oscillon,weigel2014kink,takyi2016coll,weigel2019collective,manton2021kink,dorey2021res,sutcliffe2022bound} in which one only encounters solitary waves. For most of the integrable systems one is usually in the luxurious position to be able to carry out some explicit quality check to test the success and shortcomings of the moduli space approach by comparing with the exact multi-soliton solutions. 
For non-integrable systems this possibility usually does not exist in an exact manner. However, for those systems the collective coordinate approach provides especially insightful information as it allows for the separation of different types of modes present in these kind of systems \cite{kivshar1989dynami,sugiyama1979kink,belova1988quasi,caputo1991kink,quintero2000ano,GoodHab2005,goodman2007chaotic,romanc2010oscillon,weigel2014kink,takyi2016coll,weigel2019collective,manton2021kink,dorey2021res,sutcliffe2022bound} .

Here we wish to extend the approach to integrable field theories that posses stable complex multi-soliton solutions with real energies when suitable regularized by implementing a ${\cal PT}$-symmetry \cite{CenFring,CorreaFring,fring2020BPS}, i.e. a simultaneous reflection in space and time. This technique has been adapted using ideas originally introduced in the context of non-Hermitian quantum theories \cite{Urubu,Bender:1998ke,Alirev,PTbook}.

 In principle there are various possibilities on how one might extend the moduli space approach to non-Hermitian theories. The most obvious and general option is to consider a non-Hermitian field theory from the very start, which possibly also leads to a non-Hermitian reduced system in terms of the collective coordinates. The second option is to consider a Hermitian field theory that possess complex solutions in parts of their parameter space so that it leads to a non-Hermitian reduced system. Here we will focus at first on a theory that is Hermitian as a field theory and also in its reduced variant, but possess complex solutions. The Bullough-Dodd model \cite{BDodd,zhiber1979klein} is an integrable field theory with such features and will serve here as a sample theory. We will construct its collective coordinates and compare with the exact multi-soliton solutions. 
 
 Our manuscript is organised as follows: In section 2 we recall the key ideas of the moduli space approach when applied to a 1+1 dimensional field theory. In Section 3 we carry out the moduli space analysis for the Bullough-Dodd model, we discuss the moduli spaces associated to complex one and two-soliton solutions, recall the zero modes of the model, derive analytical expressions for the collective coordinates by comparing with the exact soliton solutions, conjecture the two collective coordinates that describe a newly found triple bounce scattering and explain how to derive spatial displacements and time-delays by using the moduli. Our conclusions are stated in Section 4.  The calculation of the integrals occurring in the derivation of the reduced metric and potential are explicitly presented in an appendix. 

\section{Moduli spaces for 1+1 dimensional scalar field theories}

We start by recalling some well-known features of the usage of a moduli space analysis to 1+1 dimensional field theories, see e.g. \cite{takyi2016coll,manton2021kink} for recent clear expositions. The key idea of this approach consists of reducing the classical field theory to a classical mechanical system describing point particles in the hope that many properties of the original theory are captured in its simplified version. We consider here a scalar field theory for which the corresponding Lagrangian densities are replaced by their reduced versions 
\begin{equation}
\mathcal{L=}\frac{1}{2}\partial _{\mu }\varphi \partial ^{\mu }\varphi
-V(\varphi ) \quad \rightarrow \quad L_{\text{red}}=T-V_{\text{red}}. \label{lag}
\end{equation}%
Throughout the manuscript we adopt here the Lorentzian spacetime metric $\limfunc{diag}(1,-1)$. In principle one might allow the field $\varphi(x,t) $ and potential $V(\varphi ) $ to be complex and possibly even non-Hermitian, but we will stick  at first with a Hermitian version. Next we recall how the reduced Lagrangian $L_{\text{red}}$ is constructed. As a starting point one takes the static solutions $\phi (x,\mathbf{u})$ to the Euler-Lagrange equation for the classical field theory 
\begin{equation}
	\phi ^{\prime \prime }-\frac{\partial V(\phi )}{%
		\partial \phi }=0,  \label{EL}
\end{equation}
depending on the space coordinate $x$ and a set of constants  $\mathbf{u}=(u^{1},\ldots ,u^{n})$, i.e. we replace $\varphi(x,t) \rightarrow  \phi (x,\mathbf{u})$ with $\dot{\phi}=0$.
 The constants are interpreted as the moduli or collective coordinates spanning the moduli space of dimension $n$, which is determined by the number of linearly independent zero modes. In turn, zero modes are the zero energy solutions to the auxiliary Sturm-Liouville eigenvalue problem that arises in the stability analysis of a linear perturbation, see e.g. \cite{correa2022linearly} and references therein. In the moduli space approach the constants are subsequently elevated to dynamical variables depending explicitly on time, $\mathbf{u} \rightarrow \mathbf{u}(t)$. In terms of these variables we can now define the kinetic energy $T$ of the field evolution restricted to the moduli space and the reduced potential $V_{\text{red}}$ occurring in the reduced Lagrangian (\ref{lag})
\begin{eqnarray}
	T\!\!&=&\!\!\frac{1}{2}\int\nolimits_{-\infty }^{\infty }\dot{\phi}^{2}dx=\frac{1}{2}%
	\int\nolimits_{-\infty }^{\infty }\frac{\mathcal{\partial \phi }}{\partial
		u^{i}}\frac{\mathcal{\partial \phi }}{\partial u^{j}}\dot{u}^{i}\dot{u}%
	^{j}dx=\frac{1}{2}g_{ij}\dot{u}^{i}\dot{u}^{j},\\
  V_{\text{red}} \!\!&=&\!\!\int\nolimits_{-\infty }^{\infty }\left[ \frac{1}{2}\left( \phi ^{\prime}\right) ^{2}-V(\phi )\right] dx,  \label{Lred}
\end{eqnarray}
where, together with $\dot{\phi}=(\partial \phi /\partial u^{i})\dot{u}^{i}$, the target space metric depending on the moduli space coordinates has been put in place as
\begin{equation}
	g_{ij}(\mathbf{u})=\int\nolimits_{-\infty }^{\infty }\frac{\mathcal{\partial
			\phi }}{\partial u^{i}}\frac{\mathcal{\partial \phi }}{\partial u^{j}}dx.
	\label{metric}
\end{equation}%
The dynamics of the newly introduced $n$-point particles is now simply obtained from classical Euler-Lagrange equations
\begin{equation}
	\frac{\partial L_{\text{red}}({\bf u } , \dot{{\bf u }} )}{\partial u_i } - 	\frac{d}{dt}\frac{\partial L_{\text{red}}({\bf u } , \dot{{\bf u }} )}{\partial \dot{u}_i } =0, \qquad i=1,\ldots, n .  \label{classEL}
\end{equation}

A time-dependent solutions $\varphi(x,t)$ to the full Euler-Lagrange equation can be constructed from the static solution by a Lorentz boost $\varphi(x,t)=\phi[(x-vt)/\sqrt{1-v^2}]$, where $v$ denotes the velocity. We shall also make use of the energies for particular solutions, which are computed from energy densities $\varepsilon(\varphi)$ as
\begin{equation}
  E[\varphi]= \int\nolimits_{-\infty }^{\infty } dx  \varepsilon(\varphi),   \qquad  \varepsilon(\varphi)= \left( \frac{1}{2}\dot{\varphi}^{2}  +\frac{1}{2}  (\varphi^{\prime } )^2  +V(\varphi )  \right).   \label{ener}
\end{equation}
Here our main objective is to investigate the properties of reduced theories that result from complex solutions to the Euler-Lagrange equations that lead to theories that are apparently ill-defined due to the fact that their potentials are not bounded from below, have an indefinite or non-invertible target space metric. Moreover the solutions we consider possess singularities that in general lead to infinite energies $\varepsilon(\varphi)$, but can be regularized by choosing some constants in the solutions in such a way that they become ${\cal PT}$-regularized.  

\subsection{Solutions and quality checks for the moduli}
It appears that solutions in terms of the collective coordinates as functions of $t$ describing the dynamics of the reduced classical point particle systems are especially useful for systems that are not integrable and explicit solutions to the field equations are not available. However, even when having exact solutions at hand the moduli space formulation might be much simpler and in addition one may compare the two alternative versions in a number of ways. Implicit solutions are readily obtained \cite{takyi2016coll,manton2021kink}. Assuming for instance that we only have one collective coordinate (modulus), say $u(t)$, one can solve the energy equation
\begin{equation}
\frac{1}{2} g(u) \dot{u}^2 + V(u) = E \label{encons},
\end{equation}
in a straightforward manner to
\begin{equation}
	\int\nolimits_{u_0 }^{u(t) } \sqrt{ \frac{g(\tilde{u})}{2[E-V(\tilde{u})]} } d\tilde{u} =\pm (t-t_0) . \label{modsol2}
\end{equation}
Here one could think of the energy as $E[\varphi]$ or $E(u)$. This means we may involve either the exact solutions from the field theory $\varphi$ or are able to read off the ``exact" expression for $u(t)$ from the exact solution $\varphi$. Alternatively, by staying fully within the moduli space formulation we may attempt to solve directly the classical Euler-Lagrange equation (\ref{classEL}) of the reduced theory 
\begin{equation}
\ddot{u} =\frac{1}{g(u)} \left( \frac{\partial g}{\partial u} \dot{u}^2 + \frac{\partial V}{ \partial u}       \right), \label{ELred}
\end{equation}
with some suitable initial conditions. The result obtained from (\ref{modsol2}) or (\ref{ELred}) may then be compared with the exact solution for $u(t)$, subject to the possibility to extract it from the exact solution of the field theory.

Alternatively \cite{weigel2014kink,takyi2016coll}, when thinking of the collective coordinate as reducing the field evolution to the classical evolution of a particle, we can also extract the collective coordinate from the mean value or expectation value of $x$
\begin{equation}
              \left<  x \right>_t = \frac{\int x \varrho[\varphi] dx}{\int  \varrho[\varphi] dx} . \label{expx}
\end{equation}
Here $\varrho[\varphi]$ in (\ref{expx}) can be any density  of a conserved quantity $Q=\int  \varrho[\varphi] dx$ with $\dot{Q}=0$. For instance, the energy density $\varepsilon(\varphi)$, as introduced in (\ref{ener}), or even powers of it \cite{weigel2014kink} is most readily available \cite{takyi2016coll}. The limits of the integrals have to be suitably adjusted according to the interpretation of what the quantity $\left<  x \right>_t$ should correspond to in the moduli space picture.

\section{Moduli spaces for the Bullough-Dodd model}

We will now consider a concrete example and construct the moduli spaces for the Bullough-Dodd model \cite{BDodd,zhiber1979klein}, which
is an integrable scalar field theory defined by the Lagrangian density 
\begin{equation}
\mathcal{L=}\frac{1}{2}\partial _{\mu }\varphi \partial ^{\mu }\varphi
-e^{\varphi }-\frac{1}{2}e^{-2\varphi }+\frac{3}{2}. \label{BDLag}
\end{equation}
The  classical as well as the quantum integrability of the model has been exploited in \cite{andreevback} and \cite{Fring:1992pj}, respectively. The main aspect we will be interested in here are the consequences for the moduli space construction of the feature that some of the classical multi-soliton solutions constructed in \cite{assis2008bullough,correa2022linearly} are complex in parts of their parameter regime. Using arguments from \cite{CenFring,CorreaFring,fring2020BPS} we recently established \cite{correa2022linearly} that as an effect of the underlying ${\cal PT}$-symmetry and integrability the energies of these solutions are real, including those for the complex solutions. Furthermore, a stability analysis revealed \cite{correa2022linearly} that some of the solutions are stable whereas others unstable. Crucially for our moduli space analysis also the zero modes were constructed in \cite{correa2022linearly}.

We require here the classical Euler-Lagrange equation of motion (\ref{classEL}) for the Bullough-Dodd model (\ref{BDLag}), which is the nonlinear equation 
\begin{equation}
\ddot{\varphi}-\varphi ^{\prime \prime }+e^{\varphi }-e^{-2\varphi }=0 .
\label{BDequ}
\end{equation}%
We present and analyse now the moduli spaces for various types of soliton solutions to this equation.

\subsection{One-soliton solutions with real or purely imaginary modulus} \label{onesolrealpure}
The simplest type of solution, a one-soliton solution, 
\begin{equation}
\varphi_I^{\pm} (x,t)= \ln \left[\frac{\cosh \left(\beta +\sqrt{k^2-3} t+k x\right) \pm 2}{\cosh \left(\beta +\sqrt{k^2-3} t+k
	x\right)\mp 1}\right],  \quad \beta \in  \mathbb{C}, \label{sol1}
\end{equation}
was found to be stable in \cite{correa2022linearly}. The solution $\varphi_I^{+}$ is a bright real peakon for $\beta \in \mathbb{R}$ and $|k|> \sqrt{3}$, possessing a singularity at $x_0=-[ \beta +\sqrt{k^2-3} t]/k$. Taking therefore the constant $\beta$ to be purely imaginary, i.e. $\beta \in i \mathbb{R}$, appears to be very natural as this choice removes the singularity and makes the solution ${\cal{PT}}$-symmetric, i.e. it remains invariant under the transformation ${\cal{PT}}: x \rightarrow -x,  t \rightarrow -t,  i \rightarrow -i,  \varphi \rightarrow \varphi$. Moreover, with this choice the energy in (\ref{ener}) becomes finite, which would otherwise diverge. This process of choosing constants in the complex solutions that render it ${\cal{PT}}$-symmetric is referred to as {\it{${\cal{PT}}$-regularization}}. In turn, $\varphi_I^{-}$ is real a dark double peakon for $\beta \in \mathbb{R}$, $|k|> \sqrt{3}$ and $x<x_+$, $x>x_-$ where $x_\pm=-[\beta \pm \arccosh(2) +\sqrt{k^2-3} t]/k$. For the same reasons also this solutions needs to be ${\cal{PT}}$-regularized. More details and sample plots of these solutions can be found in \cite{correa2022linearly}. Here we will not explore the regime $|k|< \sqrt{3}$ of the parameter space when $\varphi_I^{\pm}$ become complex breather-like solutions.

Next we construct the moduli spaces corresponding to these solutions. For this purpose we have to solve the static version of the Euler-Lagrange equation (\ref{EL}) to start with. As we already have an exact time-dependent solution, the static solution is trivially obtained in this case by the limit
\begin{equation}
\lim_{\substack{k \rightarrow k_s =  \sqrt{3} \\ \!\!\! \!\!\! \!\!\!\!\!\!\!\!\!\beta \rightarrow u}}  \varphi_I^{\pm} (x,t) =	\phi_I^{\pm} (x,u)= \ln \left[\frac{\cosh \left(u +\sqrt{3} x\right) \pm 2}{\cosh \left(u + \sqrt{3}
		x\right)\mp 1}\right],  \quad u \in  \mathbb{C}. \label{sol1stat}
\end{equation}
In turn, the time-dependent solution is recovered by the aforementioned Lorentz boost with $v= \pm \sqrt{k^2-3}/k$. We identified here the constant $\beta$ as our collective coordinate $u$. At first we take $u$ to be real and elevate it to be a time-dependent function $u \rightarrow u(t)$. Using equation (\ref{metric}), we compute the target space
metric for the reduced theory as $g^{\pm}=- 2\sqrt{3}$ and by evaluating the expression in (\ref{Lred}) we
obtain the reduced potential $V_{\text{red}}^{\pm} (u)=-6 \sqrt{3}$. Therefore with (\ref{lag}) we assemble these results to the reduced Lagrangian, which immediately gives the corresponding Euler-Lagrange equation and its solution
\begin{equation}
L_{\text{red}}^{\pm}=- \sqrt{3}\dot{u}^{2} + 6 \sqrt{3}, \qquad    \overset{(2.10)}{\Rightarrow}  \quad \ddot{u}=0  \quad \Rightarrow \quad u(t) =  c_1 t + c_2 .  \label{redequn}
\end{equation}
With the particular choice of the constants $c_1 = \sqrt{k^2 -3}$, $c_2= \beta$ we recover the exact value for $u(t)$ that we may read off from (\ref{sol1}) when compared to (\ref{sol1stat}). 

A second possible choice is based on the ${\cal PT}$-symmetry argument mentioned above, so that we can also take $u \rightarrow i u $ in (\ref{sol1stat}), which gives the same result as in (\ref{redequn}), because $u$ is simply normalised with a different constant. 

As a third option one may be tempted to associated the model to a two dimensional moduli space by setting $u \rightarrow u_1+ i u_2$. However, in this case we obtain the metric
\begin{equation}
	g^{\pm}= 2 \sqrt{3} \left(
	\begin{array}{cc}
		-1 & - i \\
		- i &  1 \\
\end{array}
\right) ,
\end{equation}
which is not invertible as $\det g^{\pm}= 0 $. This is due to the fact that the moduli space is only one dimensional as confirmed in \cite{correa2022linearly}, where we only identified one zero mode which must be identical to the dimension of the moduli space. We recall this briefly in section 3.3.

 Computing the expectation values of $x$ by means of (\ref{expx})  for the one-soliton solution (\ref{sol1}) gives
\begin{equation}
	\left<  x \right>_t = -\frac{1}{k} \left( \sqrt{k^2-3} t -  \beta \right) . 
\end{equation}
We obtain the same result by using instead of $\varrho[\varphi]$ the energy density $\varepsilon[\varphi]$ or the momentum density 
$\varrho[\varphi]= {\varphi '}\dot{\varphi}$. The result is to be expected, as is most obvious for $\varphi^+$ for which the solution is well localised, so that $\left<  x \right>_t =x_0$ corresponds to the centre of mass. For $\varphi^-$ we interpret the expectation value as  $\left<  x \right>_t =(x_-+x_+)/2$, which is also its centre of mass.

\subsection{One-soliton solutions with complex moduli}
The second type of one-soliton solution 
\begin{equation}
\varphi_{II}^{\pm} (x,t)	= \ln \left\{ \omega^{\mp 2} \left[1-\frac{6  \beta  e^{t \sqrt{k^2+ 3 \omega^{\pm 1}}+k
		x}}{\left(1+\beta  e^{t \sqrt{k^2+3 \omega^{\pm 1}}+k x}\right)^2}\right]\right\} , \label{sol2}
	\end{equation}
is always complex and does not possess a regime in which the solution is real. Here $\omega=e^{i \pi/3}$ denotes the third root of unity. Sample solution are depicted in \cite{correa2022linearly}, showing that the real part of this solution is an oscillation between a regular shaped soliton solution and a double peakon solution. The solution has the property of having real energies despite being complex, but they are still uninteresting as they were found in \cite{correa2022linearly} to be unstable. Thus in principle we would like to discard them right away for this reason. We show here briefly for the one-soliton solutions that also their corresponding moduli spaces are ill-defined.

The static one-soliton solution is obtained in this case by means of the limit
\begin{equation}
	\lim_{k \rightarrow k_s = \pm \sqrt{3} \omega^2 }  \varphi_{II}^{\pm} (x,t) =	\phi_{II}^{\pm} (x,u)= \ln \left[ \omega^{\mp2}  \frac{\cos(\sqrt{ 3 }\omega^{\pm 1/2} x\pm  u ) -2}{ \cos( \sqrt{ 3 }\omega^{\pm 1/2} x\pm u ) +1}       \right],  \quad u \in  \mathbb{C}. \label{sol2stat}
\end{equation}
Since the coefficient of $t$ is now always complex it seems natural to introduce two collective coordinates on a complex moduli space, say $a \in \mathbb{R}$ and $b\in \mathbb{R}$, as
\begin{equation}
	t \sqrt{k^2+ 3 \omega^{\pm 1}} \rightarrow a+i b,
\end{equation}
such that the static solution becomes
\begin{equation}
\phi_2^{\pm} (x,a,b)	=\ln \left\{ \omega^{\mp 2} \left[ 1-\frac{6 \beta _0 \beta _1 e^{a+i b+\sqrt{3} x \omega^{\pm 2}}}{\left(\beta
	_0+\beta _1 e^{a+i b+\sqrt{3} x \omega ^{\pm 2}}\right)^2}\right] \right\} . \label{solstat2}
\end{equation}
However, for this choice the target space metrics computed by means of (\ref{metric}) result to
\begin{equation}
g^{\pm}= 2 \sqrt{3} \left(
\begin{array}{cc}
	\omega ^{\pm 4} & - \omega ^{\pm\frac{ 5}{2}} \\
	- \omega ^{\pm \frac{5}{2}} &  \omega ^{\pm 1} \\
\end{array}
\right) ,
\end{equation}
which are evidently not invertible as $\det g^\pm=0$. Setting either $a$ or $b$ to zero still leaves us with non-Hermitian kinetic energy terms
\begin{equation}
T^{\pm}(a)=\sqrt{3} \omega^{\pm 4} \dot{a}^{2},  \qquad T^{\pm}(b)=\sqrt{3} \omega^{\pm 1} \dot{b}^{2} .
\end{equation}
Dyson maps for these systems are easily found. Defining $\eta_a^{\pm} =\exp( \pm 4/3 \pi \dot{a} a )$ and 
$\eta_b^{\pm} =\exp( \mp  2/3 \pi \dot{b} b)$ we calculate the Hermitian counterparts from the adjoint action
\begin{equation}
	\eta_a^{\pm} T^{\pm}(a) \eta_a^{\mp} =\sqrt{3}\dot{a}^{2} \qquad \text{and} \qquad \eta_b^{\pm} T^{\pm}(b) \eta_b^{\mp} =\sqrt{3}\dot{b}^{2} .
\end{equation}
The reduced potential computed from (\ref{Lred}) is vanishing.

\subsection{Zero modes for the Bullough-Dodd solutions}

We recall from \cite{correa2022linearly} that the Sturm-Liouville eigenvalue problem of the stability analysis for the Bullough-Dodd model reads 
\begin{equation}
	- \Phi_{xx} +  V(x) \Phi = \omega^2 \Phi \qquad \text{with} \quad V(x)=\left( e^{\phi(x)} + 2 e^{-2 \phi(x)}  \right) . \label{BDtisch}
\end{equation}
The static solution for the cusp and oscillatory solution acquire the form
\begin{equation}
	\phi^{\pm}_1(x) = \ln \left[\frac{\cosh \left(\beta +\sqrt{3} x\right) \pm 2}{\cosh \left(\beta +\sqrt{3} x\right) \mp 1}\right], 
\end{equation}
which when substituted into (\ref{BDtisch}) yield the potentials
\begin{equation}
	V^{\pm}(x) = 1-\frac{3}{1 \mp \cosh \left(\beta +\sqrt{3} x\right)}+\frac{8 \sinh ^4\left[\frac{1}{2} \left(\beta +\sqrt{3}
		x\right)\right]}{\left[2 \pm \cosh \left(\beta +\sqrt{3} x\right) \right]^2} . \label{pot}
\end{equation}
We think now of the larger class of functions depending on the continuous parameter $u$ as $\psi^{\pm}(x,u)=\phi^{\pm}(x+u/\sqrt{3})$. The zero mode is then easily computed to
\begin{equation}
	\Phi_0^{\pm}(x) = - \frac{3 \left\{ \tanh \left[\frac{1}{2} \left(\beta +\sqrt{3} x\right)\right] \right\}^{\mp 1} }{2 \pm \cosh \left(\beta +\sqrt{3} x\right) } . \label{phizero}
\end{equation}
One verifies that the $\Phi_0^{\pm}(x)$ indeed satisfy (\ref{BDtisch}) for the potentials (\ref{pot}) with eigenvalue $\omega=0$. Notice that when taking $u \rightarrow i u$, as seems to be natural when $k < \sqrt{3}$, so that we identify  $\psi^{\pm}(x,u)=\phi^{\pm}(x+i u/\sqrt{3})$, the resulting zero mode is the same as in (\ref{phizero}) only multiplied by $i$, which simply corresponds to $u$ differently normalised. 

\subsection{Moduli spaces for two-soliton solutions}
In the construction for moduli spaces resulting from two-soliton solutions we follow the approach in \cite{manton2021kink} and start with a superposition of two one-soliton solutions involving the collective coordinate as a constant, which is of course not a solution to the equations of motion (\ref{BDequ}) due its nonlinear nature. Nonetheless, we take as our starting point 
\begin{eqnarray}
\phi^{(2)}_\pm (x,u)&=& \phi_1^{\pm} \left(k x/ \sqrt{3}+u/ \sqrt{3} \right)+\phi_1^{\pm} \left(k x/ \sqrt{3}-u/ \sqrt{3} \right) \\ \label{twosol1}
&=&  \ln \left\{1+\frac{3 \pm6 \cosh (u) \cosh (i \beta +k x)}{[\cosh (u)\mp \cosh (i \beta +k x)]^2}\right\},  \label{twosol1b}
\end{eqnarray} 
where we re-introduced a parameter $k$ with the factor $1/\sqrt{3}$ simply included for convenience. We also replaced $\beta \in \mathbb{R}$ by $i \beta$, which converts the superposition into a $\cal{PT}$-symmetric expression, i.e. it remains invariant under $\cal{PT}:$ $x\rightarrow -x$, $t\rightarrow -t$ and $i\rightarrow -i$. As is now well known \cite{CenFring,CorreaFring,fring2020BPS}, and has already been alluded to in the previous sections, this symmetry is vital for the regularization of the energy of complex solutions. In addition, we will see below that it will lead to well-defined expressions for the integrals that are required in the construction of the reduced Lagrangian which would otherwise diverge. We further notice that we may convert two solutions into each other simply by $\phi^{(2)}_- (x,u,\beta+\pi ) = \phi^{(2)}_+ (x,u, \beta)$.

The centre of the two one-solitons in (\ref{twosol1}) are located at $x_\pm = \pm u/k$, so that the modulus $u$ measures $k/2$ times the distance between the two centres. 

Let us now compute the effective Lagrangian based on the superposition of $\phi^{(2)}_+$. The metric results with (\ref{metric}) to
\begin{eqnarray}
g(u,k)&=& \int_{-\infty}^{\infty} \frac{\partial^2 \phi^{(2)}(x,u)}{\partial u^2} dx
=\int_{-\infty}^{\infty} G(x) dx \label{g2} \\
&=&  \int\limits_{-\infty}^{\infty} \frac{36 \sinh ^2 u \left[\cosh (u) \cosh (k x+i \beta )+\cosh ^2(k x+i \beta )+1\right]^2}{[2+\cosh (u-i \beta -k x)]^2 [\cosh u-\cosh
	(k x+i \beta )]^2 [2+\cosh (u+i \beta +k x)]^2} dx \notag \\ 
&=&  \frac{12}{k} \left[\frac{6 u (\cosh (2 u)-4) \coth u-4 \sqrt{3} \arccosh(2) \sinh ^2(u)}{(\cosh (2 u)-7) (\cosh (2 u)+2)}-1\right]  \notag .
\end{eqnarray} 
The integrand is here defined as the function $G(x)$ for later reference. In the appendix we explain in detail how this integral has been computed. The reduced potential resulting from the solution (\ref{twosol1}) is computed to
\begin{eqnarray}
 && V_{\text{red}}(u)= \int_{-\infty}^{\infty} \left[ \frac{1}{2} \left( \frac{\partial \phi^{(2)}}{\partial x} \right) + e^{\phi^{(2)} } + \frac{1}{2} e^{-2 \phi^{(2)} } - \frac{3}{2} \right] dx = \int_{-\infty}^{\infty} W(x) dx \label{VVred} \\
   = && \int_{-\infty}^{\infty} \left\{ \frac{18 k^2 \sinh ^2(k x+i \beta ) (1+\cosh (u) (\cosh (u)+\cosh (k x+i \beta )))^2}{[2+\cosh (u-i \beta -k x)]^2 [\cosh u-\cosh
	(k x+i \beta )]^2 [2+\cosh (u+i \beta +k x)]^2} \right. \notag \\
&&   +  \left. \frac{27 \left[1+2 \cosh u \cosh (k x+i \beta )\right]^2 \left[ 2+\left[\cosh u+\cosh (k x+i \beta )\right]^2\right]}{2 [2+\cosh (u-i \beta -k x)]^2 [\cosh u-\cosh
	(k x+i \beta )]^2 [2+\cosh (u+i \beta +k x)]^2} \right\} dx .\notag 
\end{eqnarray}
We abbreviated the integrand in (\ref{VVred}) as $W(x)$. We notice that $W(x)$  has the same pole structure, periodicity and exponentially asymptotic behaviour as the function $G(x)$ used in the calculation of the metric, so that we can use the same method and contour as explained in the appendix to compute the integral. We obtain
	\begin{eqnarray}
&& V_{\text{red}}(u)= 2 k \left[\frac{4 \sqrt{3} \arccosh(2)-u \sinh (2 u)}{\cosh (2 u)-7}+\frac{2 \left[2 u \sinh (2 u)+\sqrt{3} \arccosh(2)\right]}{\cosh
	(2 u)+2}-3\right] \qquad \label{potcomp} \\
&& +\coth u \left(\frac{45 u \text{csch}^2u}{2 k}-6 u k\right) + \frac{9}{2 k (\cosh (2 u)-7)^3} \left\{118 -291 \cosh (2 u)+30 \cosh (4 u)    \right.    \notag \\
&&  -4 \sqrt{3} \arccosh(2) [14 \cosh (2 u)+\cosh (4 u)-3
]-\cosh (6 u)+1080 \text{csch}^2(u)   \notag   \\
&& \left.+  94 u \sinh (2 u)+7 u \sinh (4 u) \right\} .\notag
\end{eqnarray}

\begin{figure}[h]
	\centering         
	\begin{minipage}[b]{0.52\textwidth}           \includegraphics[width=\textwidth]{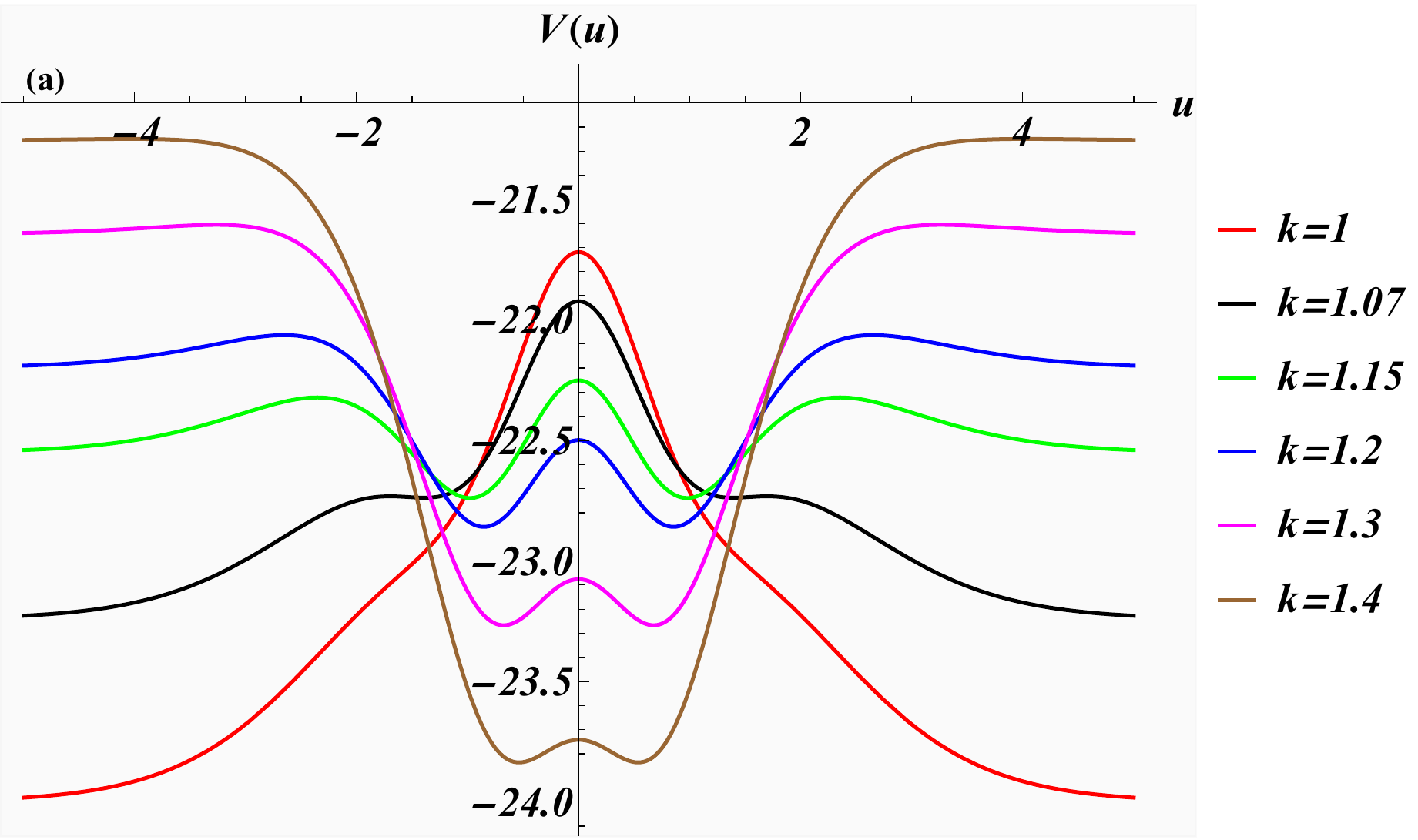}
	\end{minipage}   
	\begin{minipage}[b]{0.4\textwidth}           
		\includegraphics[width=\textwidth]{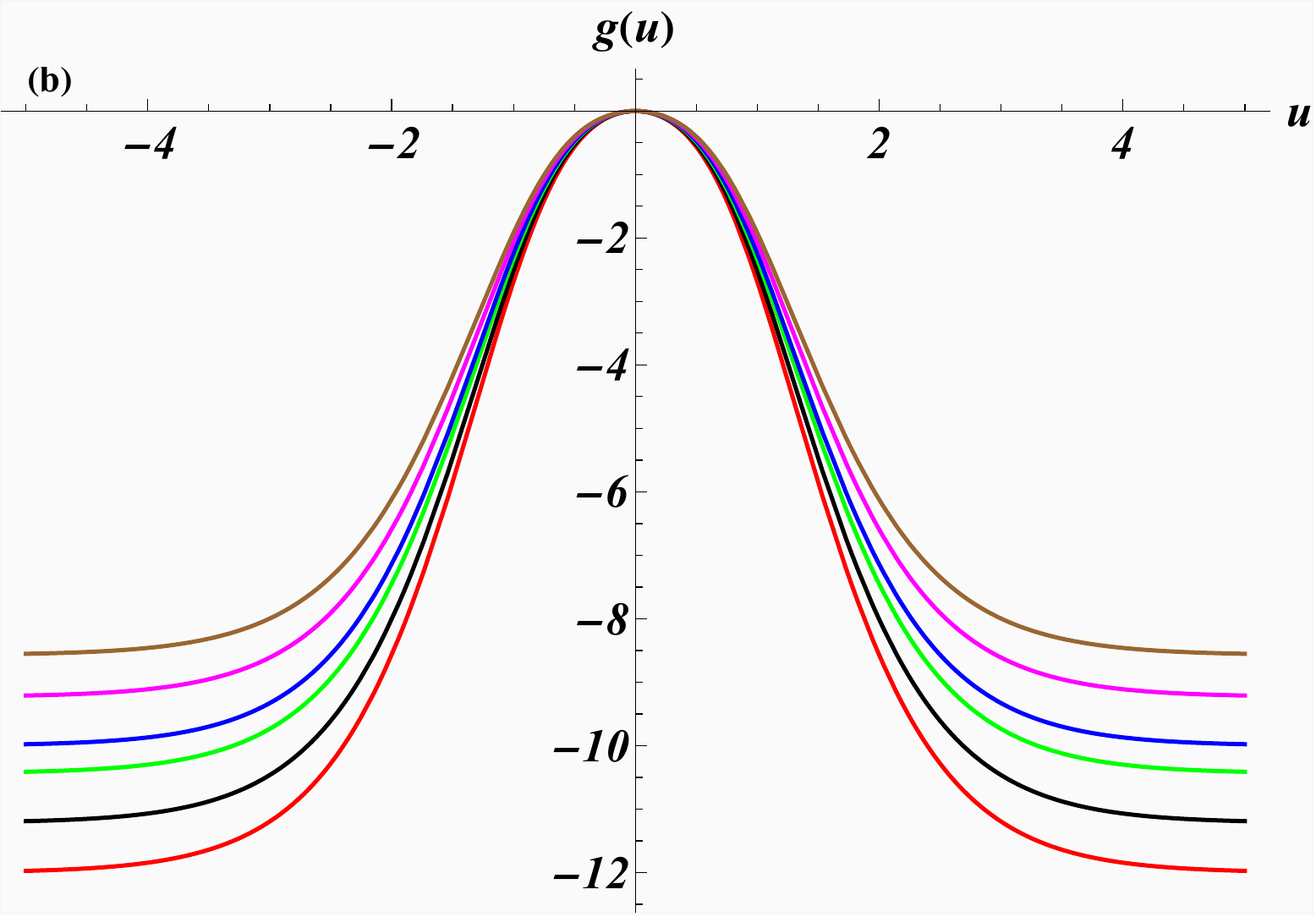}
\end{minipage}  
	\caption{Reduced potential (\ref{potcomp}) panel (a) and reduced metric (\ref{g2}) panel (b) resulting from the two-soliton solutions as functions of $u$ for different values of the parameter $k$.} 
\label{VGred2}
\end{figure}

The potential and the metric are depicted in figure \ref{VGred2} for different values of $k$. We observe that both the metric and the potential are always negative. Moreover, as we can see in panel (a) of figure \ref{VGred2} the potential changes from a scattering potential to a binding potential as $k$ increases. However, we stress that due to the negative sign in the metric the system is not a standard potential system.

\begin{figure}[h]
	\centering         
	\begin{minipage}[b]{0.8\textwidth}           
		\includegraphics[width=\textwidth]{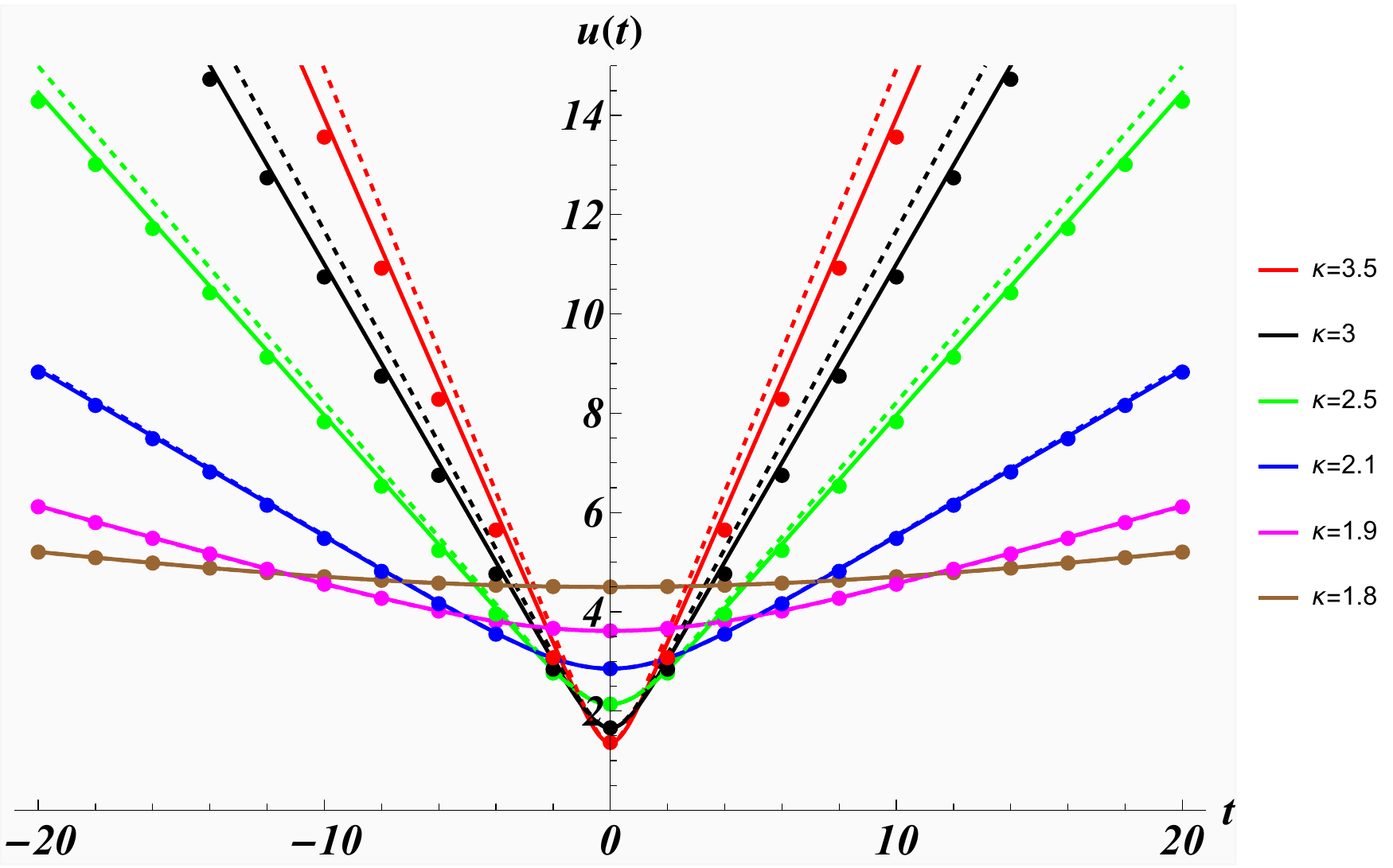}
	\end{minipage}    
	\caption{Collective coordinate $u(t)$ as solution of the Euler-Lagrange equation (\ref{ELred}) for the reduced theory (dashed lines)  versus the ``exact" expression (\ref{iden}) (solid lines) read off from the comparison of the exact two-soliton solution (\ref{exacttwo}) with a superposition of two one-soliton solutions (\ref{twosol1}) or obtained from the singularity structure of the exact solution (\ref{xmxp}) versus the average value (\ref{expx}) with $\varrho[\varphi] =\varphi $ (dots). }
	\label{compare}
\end{figure}

Having calculated the reduced metric and potential we can immediate write down the corresponding reduced Lagrangian and solve its classical Euler-Lagrange equation (\ref{ELred}). Our numerical solutions are presented in figure \ref{compare} as dashed lines. In our calculation we have taken the initial values to be $\dot{u}(0)=0$, $u(0)=\arccosh[P(\kappa)]$.  This choice will be justifies below. The same result is obtained from the integration of the conservation of energy equation (\ref{modsol2}) with an adjustment of the initial conditions from $\dot{u}(0), u(0)$ to $\dot{u}(0), E$. Needless to say that different types of initial conditions will also produce different types of qualitative behaviour. 

In figure \ref{compare} we also present the average value of the distance $2x$ as stated in (\ref{expx}). Both integrations are taken over $[0,\infty)$ and the density is simply related to the field itself, i.e. we take $\varrho[\varphi] =\varphi $. As initial conditions we take once $\dot{u}(0)=0$, $u(0)=\arccosh[P(\kappa)]$, which is achieved by adjusting $\lambda$ for varying $\kappa$ as $\lambda(1.8)=44.543$, $\lambda(1.9)=18.1828$, $\lambda(2.1)=8.2841$, $\lambda(2.5)=3.7678$, $\lambda(3.0)=2.10736$ and $\lambda(3.5)=1.41988$. We observe that the moduli space calculation captures very well the qualitative behaviour of the exact calculation. Moreover, for certain values of $\kappa$ the fit is extremely accurate.

Let us next also explain how an analytic expression for the collective coordinate $u(t)$ may be derived.

\subsection{Analytic expressions for the modulus from the exact two-soliton solution}

In order to identify an analytic expression for the modulus we have seen already for the one-soliton solution that in principle this is possible by comparing with the exact solution. For the moduli space related to the two-soliton we compare now the superposition (\ref{twosol1}) with an exact two-soliton solution. Such a solution is found for instance in 
 \cite{assis2008bullough}
\begin{equation}
\tilde{\varphi}^{(2)} (x,t)  = \ln \left( \frac{\tau_0}{\tau_1} \right), \label{exacttwo}
\end{equation}
with tau-functions defined as
\begin{eqnarray}
	\tau_0  &=& 1-4 c_1 e^{\Gamma_1}-4 c_2 e^{\Gamma_2}+c_1^2 e^{2\Gamma_1}+ c_2^2 e^{2\Gamma_2} + 8 c_1 c_2 \frac{2 \alpha_1^4 - \alpha_1^2 \alpha_2^2 + 2 \alpha_2^4}{ (\alpha_1+\alpha_2)^2 (\alpha_1^2 + \alpha_1 \alpha_2 + \alpha_2^2)} e^{\Gamma_1+\Gamma_2} \notag \\
 	&& \!\!\!\! \!\! \!\! - 4 c_1^2 c_2 \frac{(\alpha_1-\alpha_2)^2 (\alpha_1^2 - \alpha_1 \alpha_2 + \alpha_2^2)}{ (\alpha_1+\alpha_2)^2 (\alpha_1^2 + \alpha_1 \alpha_2 + \alpha_2^2)} e^{2\Gamma_1+\Gamma_2} -  4 c_1 c_2^2 \frac{(\alpha_1-\alpha_2)^2 (\alpha_1^2 - \alpha_1 \alpha_2 + \alpha_2^2)}{ (\alpha_1+\alpha_2)^2 (\alpha_1^2 + \alpha_1 \alpha_2 + \alpha_2^2)} e^{\Gamma_1+2\Gamma_2} \notag \\
	&& \!\!\!\! \!\!  \!\! + c_1^2 c_2^2 \frac{(\alpha_1-\alpha_2)^4 (\alpha_1^2 - \alpha_1 \alpha_2 + \alpha_2^2)^2}{ (\alpha_1+\alpha_2)^4 (\alpha_1^2 + \alpha_1 \alpha_2 + \alpha_2^2)^2} e^{2\Gamma_1+2\Gamma_2} 
\end{eqnarray}
and 
\begin{eqnarray}
	\tau_1  &=& 1+2 c_1 e^{\Gamma_1}+2 c_2 e^{\Gamma_2}+c_1^2 e^{2\Gamma_1}+ c_2^2 e^{2\Gamma_2} + 4 c_1 c_2 \frac{\alpha_1^4 + 4 \alpha_1^2 \alpha_2^2 +  \alpha_2^4}{ (\alpha_1+\alpha_2)^2 (\alpha_1^2 + \alpha_1 \alpha_2 + \alpha_2^2)} e^{\Gamma_1+\Gamma_2} \notag  \\
	&& \!\!\!\!  \!\! \!\! +2 c_1^2 c_2 \frac{(\alpha_1-\alpha_2)^2 (\alpha_1^2 - \alpha_1 \alpha_2 + \alpha_2^2)}{ (\alpha_1+\alpha_2)^2 (\alpha_1^2 + \alpha_1 \alpha_2 + \alpha_2^2)} e^{2\Gamma_1+\Gamma_2} +2 c_1 c_2^2 \frac{(\alpha_1-\alpha_2)^2 (\alpha_1^2 - \alpha_1 \alpha_2 + \alpha_2^2)}{ (\alpha_1+\alpha_2)^2 (\alpha_1^2 + \alpha_1 \alpha_2 + \alpha_2^2)} e^{\Gamma_1+2\Gamma_2} \notag \\
	&& \!\! \!\!  \!\! \!\! + c_1^2 c_2^2 \frac{(\alpha_1-\alpha_2)^4 (\alpha_1^2 - \alpha_1 \alpha_2 + \alpha_2^2)^2}{ (\alpha_1+\alpha_2)^4 (\alpha_1^2 + \alpha_1 \alpha_2 + \alpha_2^2)^2} e^{2\Gamma_1+2\Gamma_2} , 
\end{eqnarray}
where $\Gamma_i = \sqrt{3} [(\alpha_i -\alpha_i^{-1})t-(\alpha_i +\alpha_i^{-1})x ]/2 $. 

\begin{figure}[h]
	\centering         
	\begin{minipage}[b]{0.45\textwidth}           
		\includegraphics[width=\textwidth]{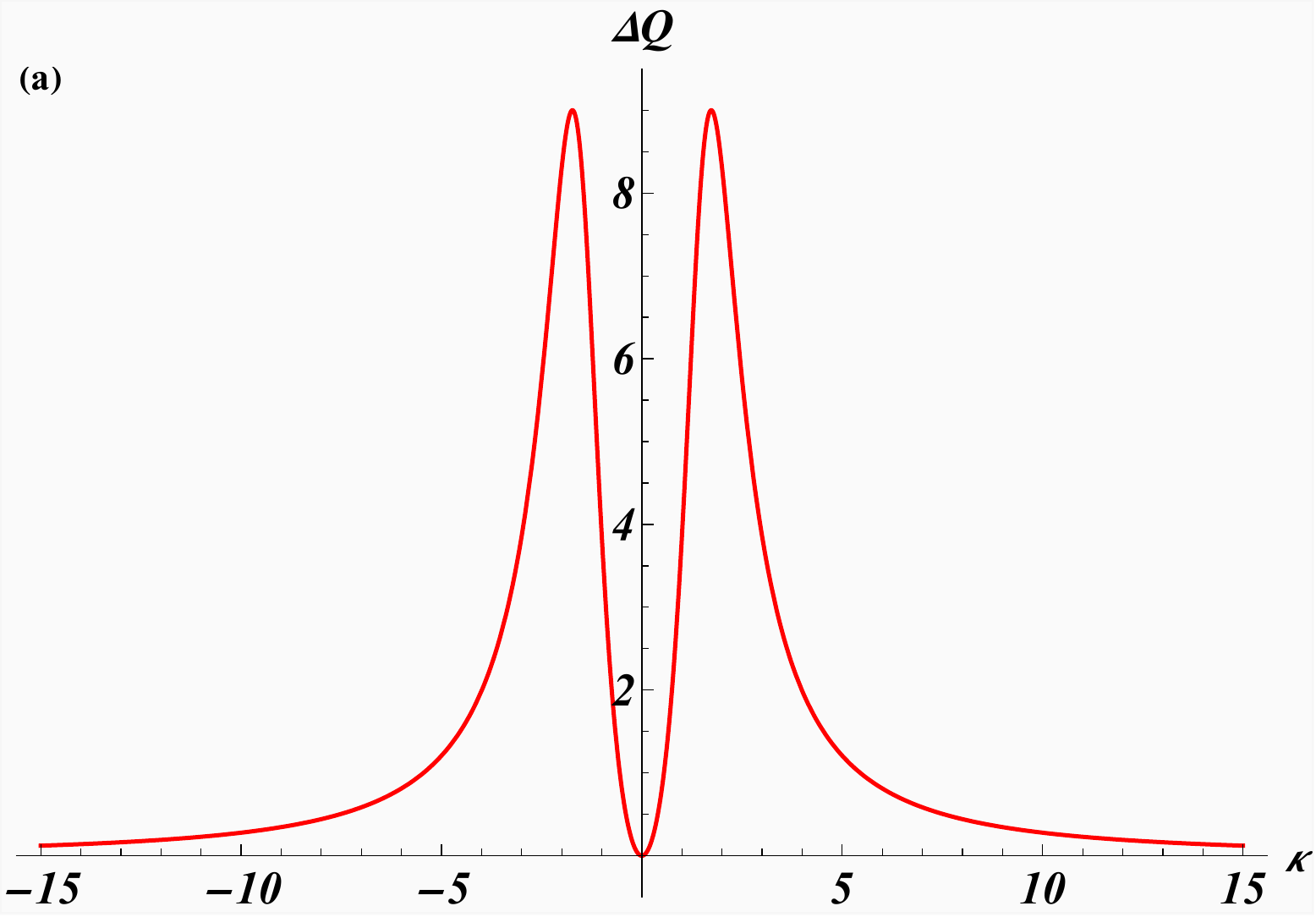}
	\end{minipage}  
	\begin{minipage}[b]{0.54\textwidth}           
		\includegraphics[width=\textwidth]{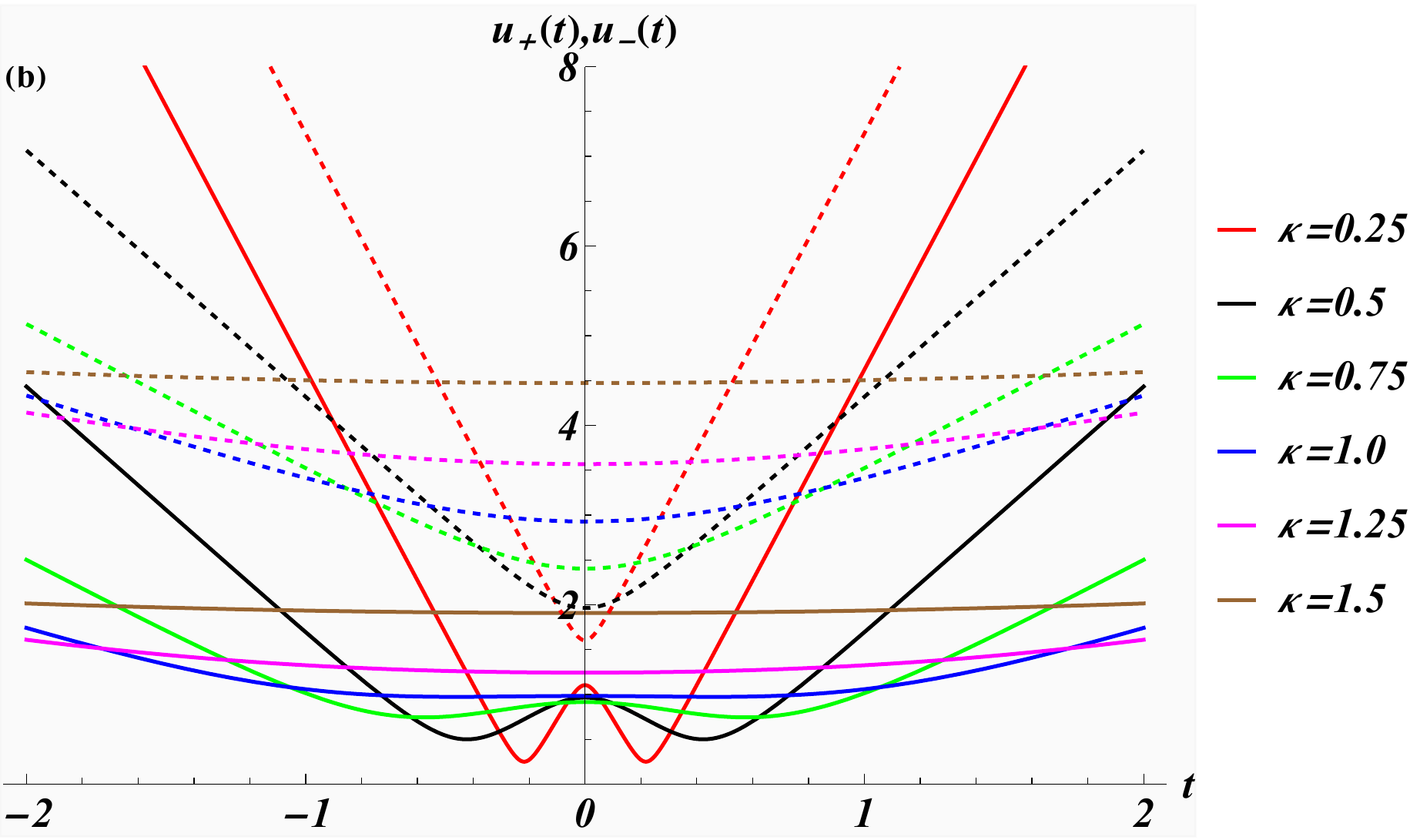}
	\end{minipage}   
	\caption{Panel (a): Deviation $\Delta Q(\kappa) = Q(\kappa) -Q_e$ as a functions of $\kappa$ from the exact value $Q_e$. Panel (b): Exact moduli $u_-(t)$ (solid lines) and  $u_+(t)$ (dashed lines) from (\ref{xmxpnew}) exhibiting triple bouncing for some values of $\kappa$.}
	\label{Gkappafig}
\end{figure}

In order to match with the superposition (\ref{twosol1}) we have to reduce the amount of free parameters from two, $\alpha_1$, $\alpha_2$, to one, say $\kappa$. This process is akin to constructing a degenerate multi-soliton solution from a non-degenerate one, see \cite{CorreaFring,CCFsineG}. In fact, in \cite{manton2021kink} a similar comparison was carried out for a two-kink superposition in the sine-Gordon model, where the authors compared, see expression in equation (33), precisely with the exact degenerate two-kink solution constructed in \cite{CCFsineG}, see equation (2.12) therein. Here we make the 
specific choices for the parameters in the solution $\alpha_1 \rightarrow \sqrt{3}/ \kappa $, $\alpha_2 \rightarrow \kappa / \sqrt{3} $, $c_1 \rightarrow - \lambda$, $c_2 \rightarrow - \lambda$, such that $\tilde{\varphi}^{(2)}$ acquires the form 
\begin{equation}
	 \tilde{\varphi}^{(2)} (\kappa,x,t)  = \ln \left( 1 + \frac{Q(\kappa) + 6 P(\kappa) \cosh\left[ \left(\frac{3}{2 \kappa}- \frac{\kappa}{2} \right) t  \right] \cosh\left[ \left(\frac{3}{2 \kappa}+ \frac{\kappa}{2} \right) x + \beta(\kappa)  \right] } {\left(P(\kappa) \cosh\left[ \left(\frac{3}{2 \kappa}- \frac{\kappa}{2} \right) t  \right]-  \cosh\left[ \left(\frac{3}{2 \kappa}+ \frac{\kappa}{2} \right) x + \beta(\kappa)  \right] \right)^2} \right),  \label{twosolsp}
\end{equation}
where
\begin{equation}
    P(\kappa) := \frac{(\kappa^2 +3) \sqrt{\kappa^4 + 3 \kappa^2 +9}}{(\kappa^2 -3) \sqrt{\kappa^4 - 3 \kappa^2 +9}}, \quad \beta(\kappa):= \frac{1}{2} \ln\left( \frac{P^2(\kappa)}{\lambda^2}\right), \quad Q(\kappa):= \frac{3(\kappa^2 +3)^2}{\kappa^4 - 3 \kappa^2 +9}.
\end{equation}
Comparing now (\ref{twosolsp}) with (\ref{twosol1b}) we notice that the two solutions formally coincide, i.e. $ \tilde{\varphi}^{(2)} (x,t) \rightarrow \varphi^{(2)} (x,u)$ for 
\begin{equation}
  Q(\kappa) \rightarrow Q_e=3, \qquad \frac{3}{2 \kappa}+ \frac{\kappa}{2} \equiv k, \qquad u(t) \equiv \arccosh \left\{P(\kappa) \cosh\left[ \left(\frac{3}{2 \kappa}- \frac{\kappa}{2} \right) t  \right] \right\} . \label{iden}
\end{equation}
While the last two relations in (\ref{iden}) are exact identifications, the first relation can only be achieved at the ill-defined values $\kappa=0, \kappa \rightarrow \pm \infty $. Panel (a) in figure \ref{Gkappafig} shows the deviation of $Q(\kappa)$ from the exact value for all values of $\kappa$.  

\begin{figure}[h]
	\centering         
	\begin{minipage}[b]{0.49\textwidth}           
		\includegraphics[width=\textwidth]{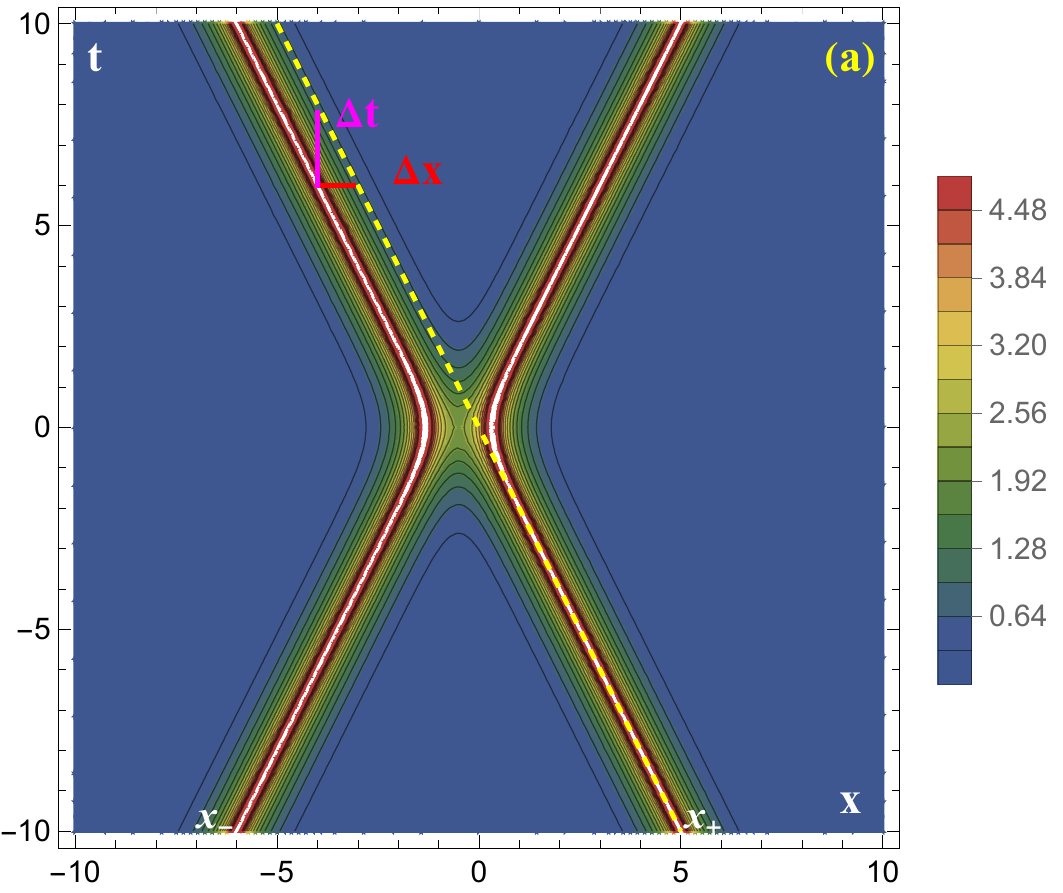}
	\end{minipage}  
	\begin{minipage}[b]{0.49\textwidth}           
		\includegraphics[width=\textwidth]{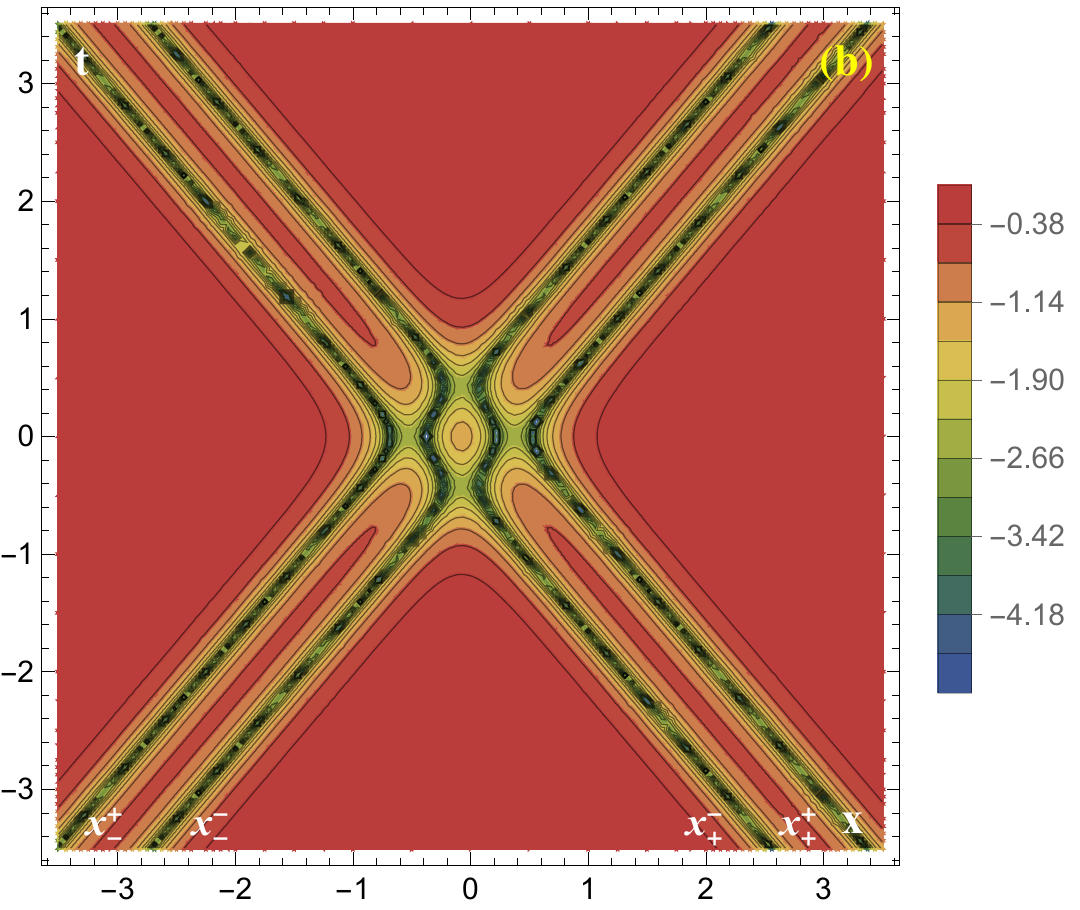}
	\end{minipage}   
	\caption{Panel (a): Non-regularized bright two-peakon scattering for $\kappa=3$, $\lambda=1$ together with a bright one peakon (dashed yellow line). The one-peakon constituent within the two-peakon solution is tracked by a one-peakon solution with the same speed from the infinite past to the infinite future. The deviation between the two pathes in the future in space and time is the spatial displacement $\Delta x$ and time-delay $\Delta t$, respectively. The poles $x_\pm$ as identified in (\ref{xpmpole}) are indicated. Panel (b): 
	non-regularized dark two-double peakons scattering for $\kappa=0.5$, $\lambda=1$ exhibiting a triple bouncing.
		The poles $x_\pm^\pm$ as identified in (\ref{xpmpoledark}) are indicated.}
	\label{DarkBright}
\end{figure}

We explore here the entire regime for $\kappa \in \mathbb{R}$, noting that for $\kappa > \sqrt{3}$ we have a scattering between two bright peakons, panel (a) in figure \ref{DarkBright}, whereas for $\kappa < \sqrt{3}$ we obtain a scattering between two dark double peakons as depicted in panel (b) of figure \ref{DarkBright}. Notice that the identification between $k$ and $\kappa$ as stated in (\ref{iden}) prevents to enter the breather region of the one-soliton solutions $\vert k \vert < \sqrt{3}$.

In figure \ref{compare} we also present the plots for the analytic expressions of (\ref{iden}). In all our numerical computations we take the values of $u(0)$ and $\dot{u}(0)$ corresponding to (\ref{iden}) as initial conditions. Notice that crossing from the bright to the dark regime is caused by the change of $P(\kappa) \rightarrow -P(\kappa)$, which is equivalent to changing $\beta(\kappa) \rightarrow \beta(\kappa) + i \pi $. We have seen that for the superposition this shift relates the $\phi^{(2)}_+ (x,u)$ to $\phi^{(2)}_- (x,u)$.   

Another possibility to obtain analytic expressions for the modulus $u(t)$  is to track directly the center of mass in the exact solution. We note that the exact unregularised two-soliton solution has poles, corresponding for $\kappa > \sqrt{3}$ to the peaks of the two one-peakon constituents, at the positions 
\begin{equation}
x_\pm= \frac{2 \kappa  }{\kappa ^2+3} \left\{ \pm \arccosh\left[ P(\kappa ) \cosh \left(\frac{\kappa ^2-3 }{2 \kappa } t \right) \right]-\beta (\kappa
)\right\}. \label{xpmpole}
\end{equation}
Thus, together with the relation between $\kappa$ and $k$, and $u(t)$ as stated in (\ref{iden}) we obtain for the distance between the two points
\begin{equation}
x_+ -x_- = \frac{4 \kappa}{3 + \kappa^2} u(t) = \frac{2}{k} u(t) . \label{xmxp}
\end{equation}
This means the function $u(t)$ obtained directly from the exact solution is identical to the one obtained from the comparison with the superposition in the muduli space analysis.

\subsection{Conjectured collective coordinates for triple bouncing}
We notice that the collective coordinate $u(t)$ for the bright and dark peakon scattering are the same. However, we clearly notice that the double peakon scattering in figure \ref{DarkBright} panel (b) exhibits a more intricate behaviour that is not captured by $u(t)$. We observe that at first the right component of the left double peakon bounces off the left component of the right double peakon. Subsequently an internal scattering process takes place in which the two components of the left and right double peakon bounce off each other. The last event in this scattering process consists of a repetition of the first scattering event, i.e. right component of the left double peakon bounces off the left component of the right double peakon. This triple bouncing appears to be a new type of scattering behaviour different from the possibilities previously discussed \cite{anco2011,cen2019asymptotic}, which were merge-split, bounce-exchange and absorb-emit scattering.   

We may therefore attempt to find a different type of moduli space that capture this triple bounce behaviour. Inspired by the success of (\ref{xmxp}) we define once more a collective coordinate by taken the difference between the singularities in the exact solution. We note first that the four singularities of the dark double peakon are located at  
\begin{equation}
	x_\pm^\pm=   \pm \frac{ 1}{k} \arccosh\left[-2 P(\kappa ) \cosh \left(\frac{\kappa ^2-3 }{2 \kappa } t \right) \pm  \sqrt{3 P(\kappa )^2 \cosh^2 \left(\frac{\kappa ^2-3 }{2 \kappa } t \right)-Q (\kappa )}      \right]-\frac{\beta (\kappa
	)}{k}, \label{xpmpoledark}
\end{equation}
where the signs in the subscript and superscript refer to the overall sign and the sign in front of the square root, respectively. We keep the relation between $k$ and $\kappa$ as stated in (\ref{iden}). See figure \ref{DarkBright} for the relative location of these poles. In analogy to (\ref{xmxp}) we therefore defined the new quantities
\begin{eqnarray}
    u_\pm(t) &=&  \frac{k}{2}\left(	x_+^\pm - x_-^\pm \right) \label{xmxpnew} \\
    &=& \arccosh\left[ \pm \sqrt{3 P(\kappa )^2 \cosh ^2\left(t\sqrt{k^2-3} \right)-Q(\kappa )}-2 P(\kappa ) \cosh \left(t \sqrt{k^2-3} \right)\right],  \notag
\end{eqnarray}
to capture the triple bouncing observed in figure \ref{DarkBright} panel (b). Indeed in figure \ref{Gkappafig} panel (b) we see that while  $u_+(t)$ exhibits the previously encountered single bouncing effect, $u_-(t)$ reproduced well the triple bouncing effect, especially for smaller values of $\kappa$. For larger values of $\kappa$ this effect seems to be smeared out. We leave it here as an open question on how these quantities may be obtained from a pure moduli space analysis involving a two dimensional reduced Lagrangian for the variables $u_-(t)$ and $u_+(t)$.    

\subsection{Spatial displacements and time-delays}

The defining feature of soliton scattering is that its one-soliton constituents 
regain their original shape after the scattering event, but are displaced in space by an amount $\Delta x$ or equivalently delayed in time by $\Delta t$, see \cite{cen2017time} and references therein for a recent exposition of systems we consider here.  From the exact solutions one may compute these quantities by tracking the positions of the singularities in the one and two-peakon solutions and compare their asymptotic values. We recall that the singularity for exact the bright one-peakon solution $\varphi_I^{+} (x,t)$ in (\ref{sol1}) is located at
\begin{equation}
	x_0(t) = - \frac{1}{k} \left(  \beta +  \sqrt{k^2 -3} \, t\right) .
\end{equation} 
From the exact expressions for the singularities in the bright two-peakon solutions in (\ref{xpmpole}), together with the identity $\arccosh (x) \simeq \ln(2 x)$ for $x \rightarrow \pm \infty$, we find the asymptotic expressions
\begin{equation}
	x_+(t) \simeq \pm \frac{1}{k} \sqrt{k^2 - 3}\, t,  \qquad 	x_-(t) \simeq \mp \frac{1}{k} \sqrt{k^2 - 3} \, t -  \frac{2}{k} \beta(\kappa),\qquad \text{for} \quad t \rightarrow \pm \infty,
\end{equation}
where we used that $\ln P(\kappa) = \beta(\kappa)$ for $\lambda=1$ and kept term of first and zeroth order in t. The locations of the singularities can then be matched asymptotically 
\begin{eqnarray}
	x_0(t) &\simeq& x_+(t),  \qquad \qquad  \qquad \text{for} \quad t\rightarrow -\infty , \label{asy1}\\
	x_0(t) &\simeq& x_-(t) + \frac{2}{k} \beta ( \kappa ),    \qquad \text{for} \quad t\rightarrow \infty .  \label{asy2}
\end{eqnarray}	
Thus tracking the one-peakon constituents within the two-peakon solution and comparing the asymptotic past (\ref{asy1}) and future (\ref{asy2}) we obtain the displacement and time-delay
\begin{equation}
\Delta x = \frac{2}{k} \beta ( \kappa ), \qquad \text{and} \qquad \Delta t = \frac{2}{\sqrt{k^2 - 3}} \beta ( \kappa ),
\end{equation}
respectively. In figure \ref{DarkBright} this comparison is graphically illustrated and explained in the caption. This interpretation presumes that we do not interpret the scattering process as back bouncing, but assume that the two solitons have exchanged their position.

Let us now see how these quantities can also be obtained from the moduli space. Recalling from section 3.1 the moduli space solution corresponding to the one-peakon, we have left and right moving solutions trough the origin   
\begin{equation}
	u_1^\pm = \pm \frac{1}{k} \sqrt{k^2 - 3} \, t.
\end{equation}
Thus, in terms of quantities in the moduli space the displacement results to
\begin{equation}
\Delta x \simeq \frac{2}{k} u_2(t) \mp \left[ 	u_1^+(t) -  	u_1^-(t)  \right], \qquad \text{for} \quad t \rightarrow \pm \infty  ,
\end{equation}
where $u_2(t)$ denotes the function $u(t)$ as constructed in (\ref{iden}). Of course without the comparison with the exact solutions the constant of integration in the moduli space can not be fixed exactly.

\section{Conclusions}

For the explicit example of the Bullough-Dodd model we have demonstrated that the moduli space formulation captures extremely well the centre of mass motion of the exact one and two-soliton solutions. However, the constructed collective coordinate $u(t)$ cannot distinguish between the bright and dark solutions as their centre of mass is identical. We conjectured two collective coordinates $u_\pm(t)$ that capture the more detailed features of the dark double double peakon solution, in particular the newly observed triple bouncing. While multiple bouncing is well-known to occur in scattering processes of non-integrable system, for integrable systems it appears to be hitherto unobserved and enlarges the previous possibilities of merge-split, bounce-exchange and absorb-emit scattering \cite{anco2011,cen2019asymptotic}. We leave the task to derive $u_\pm(t)$ from within a collective coordinates approach via the use of a reduced Lagrangian for future work. In addition, we may extract the spatial displacement and time-delay between a one-peakon constituent within the two-peakon solution and a one-peakon moving with the same speed by comparing the different moduli spaces constructed from the corresponding solutions in the same manner as in the exact case. 

In our analysis we have frequently used the ill-defined peakon expressions, solely for the reason that their singularities are easily identified. However, we stress once more that these solutions only make sense when they are suitable ${\cal PT}$-regularised as then their energy becomes finite. Moreover, the integrals that define the reduced metric and potential become also suitable regularised in this way and can be computed as explained in detail in the appendix. 

For the moduli space resulting from the one-soliton solution we have seen that the reduced metric becomes ill-defined, that is it becomes non-invertible, when we set up a moduli space with the wrong dimension or when we construct it from an unstable solution. 

We have seen that by comparing the modular spaces corresponding to the one and two soliton one can extract the spatial displacement or time-delay. However, these expressions remain generic constants that can not be fixed from within a pure moduli space analysis.

There are a number of issues that would be interesting to investigate further, a construction of moduli space that resolves the details of the dark double-peakon solution and mimic the triple bouncing, a treatment of reduced Lagrangians that correspond to non-Hermitian theories and the construction of moduli spaces associated to fully non-Hermitian field theories.

\medskip

\noindent \textbf{Acknowledgments:} FC was partially supported by Fondecyt grant 1211356 and would like to thank the Department of Mathematics at City, University of London for kind hospitality. TT is supported by JSPS KAKENHI Grant Number JP22J01230.

\begin{appendices}
	\section{The integrals for the reduced metric and potential}
We explain here in detail how the integrals for the reduced metric and potential in (\ref{g2}) and (\ref{VVred}), respectively, are computed using a method that adapts contour integrals for periodic functions. In the context of studying moduli spaces this method was advocated in \cite{pereira2021some} to evaluate integrals of a similar type for a different system. 

First of all we notice that the function $G(x)$ as defined in (\ref{g2}) is periodic and  exponentially vanishing asymptotically
\begin{equation}
G(x) = G \left(x + \frac{2 \pi i}{k} \right),  \qquad \text{and} \qquad  \lim_{x \rightarrow \pm \infty} G(x)\sim 144 \sinh^2 (u) e^{ - 2 |k x | \pm i 2 \beta   } . \label{prop}
\end{equation}
Defining a new function $\tilde{G}(x) := x G(x)$ we use the periodicity in (\ref{prop}) to observe
\begin{equation}
	\tilde{G} \left( x- \frac{i \pi}{k}   \right) -  \tilde{G} \left( x+ \frac{i \pi}{k}\right) = - \frac{2 \pi i}{k} G(x)   ,
\end{equation}
so that the integral we are interested in can we re-written as
\begin{equation}
	\int_{-\infty}^{\infty} G(x) dx = \frac{i k}{2 \pi} \lim_{R \rightarrow \infty} \left[  \int_{-R}^{R} 	\tilde{G} \left( x- \frac{i \pi}{k}   \right) - \int_{-R}^{R} 	\tilde{G} \left( x+ \frac{i \pi}{k}   \right) \right] . \label{GG}
\end{equation}

\begin{figure}[h]
	\centering         
	\begin{minipage}[b]{0.9\textwidth}           
		\includegraphics[width=\textwidth]{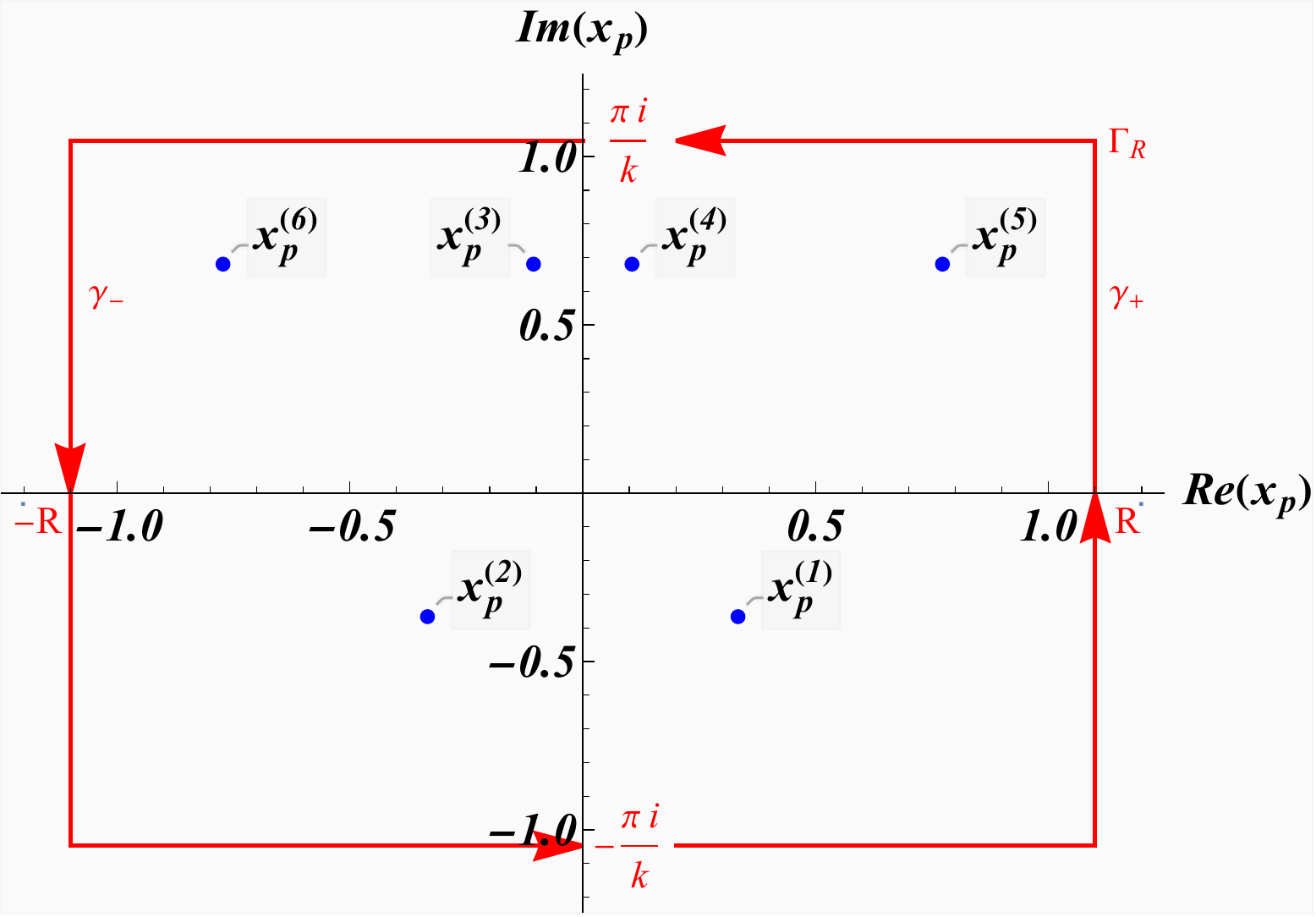}
	\end{minipage}    
	\caption{Integration contour $\Gamma_R$ and poles of $G(x)$ for $u=6/5$, $k=3$, $\beta=9/10$. }
	\label{path}
\end{figure}

Adding now two integrals along the paths $ \gamma_{\pm} = i t \pm R $ with $R \in [\mp \pi/k, \pm \pi/k]$ as indicated in figure \ref{path} to close the contour $\Gamma_R$
\begin{equation}
	\int_{\gamma_{\pm}} \tilde{G}(z) dz = \int_{\mp \pi/ k}^{\pm \pi/k} dt \frac{d \gamma_{\pm}}{dt} \tilde{G}(\gamma_{\pm}) = i \int_{\mp \pi/ k}^{\pm \pi/k} dt \tilde{G}( i t \pm R) ,   \label{addzero}
\end{equation}
we can re-express the integral in (\ref{GG}) as a contour integral of $\tilde{G}$  
\begin{eqnarray}
 	\oint_{\Gamma_{R \rightarrow \infty}} \!\! \!\!\!\!\!\! \tilde{G}(z) dz &=&  \lim\limits_{R \rightarrow \infty} \left[  \int\limits_{-R}^{R} 	\tilde{G} \left( x- \frac{i \pi}{k}   \right) +
  \int\limits_{\gamma_{+}} \tilde{G}(z) dz
  + \int\limits_{-R}^{R} 	\tilde{G} \left( x+ \frac{i \pi}{k} \right) +\int_{\gamma_{-}} \tilde{G}(z) dz  \right] \notag\\
  &=& 2 \pi i \sum\limits_{\text{Res} \Gamma_R} \tilde{G}(z) = - \frac{1}{k} 	\int_{-\infty}^{\infty} G(x) dx .
\end{eqnarray}
We have used here the fact that the two integrals, that were added in (\ref{addzero}) in order to close the contour $\Gamma_R$, are both vanishing due to the asymptotic behaviour of $G(x)$ when $R \rightarrow \infty$ as stated in (\ref{prop}).

Let us now specify this general set up for the concrete functions we are trying to integrate. Identifying at first the singularities, we see that $G(x)$ has double poles at
\begin{eqnarray}
     x_p^{(1)}&=&\frac{u-i \beta }{k}+\frac{2 i \pi  n}{k}, \quad
     x_p^{(2)}=\frac{-u-i \beta }{k}+\frac{2 i \pi  n}{k}, \label{poles1} \\  
     x_p^{(3)}&=&\frac{u-i \beta +i \pi -\arccosh2+2 i \pi  n}{k}, \quad
     x_p^{(4)}= \frac{-u-i \beta +i \pi +\arccosh2+2 i \pi  n}{k},  \\  
     x_p^{(5)}&=& \frac{u-i \beta +i \pi +\arccosh2+2 i \pi  n}{k}, \quad
     x_p^{(6)}= \frac{-u-i \beta +i \pi -\arccosh2+2 i \pi  n}{k}, \qquad \quad
     \label{poles3}
\end{eqnarray}
with $n \in \mathbb{Z}$. We only need to consider the poles in one $2 \pi i /k$-period. A sample for $n=0$ is depicted in figure \ref{path}. Notice that for real $\beta$ the poles would lie on the integration contour. In \cite{pereira2021some} this problem was overcome by introducing new variables. Here this issue is automatically resolved by having taken $\beta$ to be purely imaginary in order to obtain $\cal{PT}$-symmetric solutions. 
Summing up the residues related to the six poles for $n=0$ in (\ref{poles1})-(\ref{poles3}) leads to the expression reported in (\ref{g2}).

For the computation of the integral in (\ref{VVred}) we first notice that $W(x)$ has the same periodicity as $G(x)$ and is also  exponentially vanishing asymptotically
\begin{equation}
	W(x) = W \left(x + \frac{2 \pi i}{k} \right),  \qquad \text{and} \qquad  \lim_{x \rightarrow \pm \infty} W(x)\sim  72( k^2+3 ) \cosh^2 (u) e^{ - 2 |k x | \pm i 2 \beta   } . \label{prop2}
\end{equation}
Moreover, $W(x)$ has the same pole structure as $G(x)$ so that we can use the same contour as in figure \ref{path}. Adding up the 6 residues for $W(x)$ leads to the expression reported in (\ref{VVred}).
\end{appendices}

\newif\ifabfull\abfulltrue


\begin{thebibliography}{10}
	
	\bibitem{seiberg1988observations}
	N.~Seiberg,
	\newblock Observations on the moduli space of superconformal field theories,
	\newblock Nucl. Phys. B {\bf 303}, 286--304 (1988).
	
	\bibitem{hanany1995quantum}
	A.~Hanany and Y.~Oz,
	\newblock On the quantum moduli space of vacua of N= 2 supersymmetric SU (Nc)
	gauge theories,
	\newblock Nucl. Phys. B {\bf 452}, 283--312 (1995).
	
	\bibitem{argyres1996moduli}
	P.~C. Argyres, M.~R. Plesser, and N.~Seiberg,
	\newblock The moduli space of vacua of N= 2 SUSY QCD and duality in N= 1 SUSY
	QCD,
	\newblock Nucl. Phys. B {\bf 471}, 159--194 (1996).
	
	\bibitem{seiberg1994exact}
	N.~Seiberg,
	\newblock Exact results on the space of vacua of four-dimensional SUSY gauge
	theories,
	\newblock Phys. Rev. D {\bf 49}, 6857 (1994).
	
	\bibitem{takyi2016coll}
	I.~Takyi and H.~Weigel,
	\newblock Collective coordinates in one-dimensional soliton models revisited,
	\newblock Phys. Rev. D {\bf 94}, 085008 (2016).
	
	\bibitem{manton2021kink}
	N.~S. Manton, K.~Ole{\'s}, T.~Roma{\'n}czukiewicz, and A.~Wereszczy{\'n}ski,
	\newblock Kink moduli spaces: Collective coordinates reconsidered,
	\newblock Phys. Rev. D {\bf 103}, 025024 (2021).
	
	\bibitem{Assis:2009gt}
	P.~E.~G. Assis and A.~Fring,
	\newblock From real fields to complex Calogero particles,
	\newblock J. Phys. A: Math. Theor. {\bf 42}, 425206 (2009).
	
	\bibitem{caputo1991kink}
	J.~G. Caputo and N.~Flytzanis,
	\newblock Kink-antikink collisions in sine-Gordon and $\varphi^4$ models:
	Problems in the variational approach,
	\newblock Phys. Rev. A {\bf 44}, 6219 (1991).
	
	\bibitem{kivshar1989dynami}
	Y.~S. Kivshar and B.~A. Malomed,
	\newblock Dynamics of solitons in nearly integrable systems,
	\newblock Rev. of Mod. Phys. {\bf 61}, 763 (1989).
	
	\bibitem{sugiyama1979kink}
	T.~Sugiyama,
	\newblock Kink-antikink collisions in the two-dimensional $\varphi^4$ model,
	\newblock Prog. of Theor. Phys. {\bf 61}, 1550--1563 (1979).
	
	\bibitem{belova1988quasi}
	T.~Belova and A.~E. Kudryavtsev,
	\newblock Quasi-periodic orbits in the scalar classical $\lambda \phi^4$ field
	theory,
	\newblock Physica D: Nonlin. Phen. {\bf 32}, 18--26 (1988).
	
	\bibitem{quintero2000ano}
	N.~R. Quintero, A.~S{\'a}nchez, and F.~G. Mertens,
	\newblock Anomalous resonance phenomena of solitary waves with internal modes,
	\newblock Phys. Rev. Lett. {\bf 84}, 871 (2000).
	
	\bibitem{GoodHab2005}
	R.~H. Goodman and R.~Haberman,
	\newblock Kink-Antikink Collisions in the $\phi^4$ Equation: The n-Bounce
	Resonance and the Separatrix Map,
	\newblock SIAM J. Appl. Dyn. Sys. {\bf 4}, 1195--1228
	(2005).
	
	\bibitem{goodman2007chaotic}
	R.~H. Goodman and R.~Haberman,
	\newblock Chaotic scattering and the n-bounce resonance in solitary-wave
	interactions,
	\newblock Phys. Rev. Lett. {\bf 98}, 104103 (2007).
	
	\bibitem{romanc2010oscillon}
	T.~Romanczukiewicz and Y.~Shnir,
	\newblock Oscillon resonances and creation of kinks in particle collisions,
	\newblock Phys. Rev. Lett. {\bf 105}, 081601 (2010).
	
	\bibitem{weigel2014kink}
	H.~Weigel,
	\newblock Kink--Antikink Scattering in $\varphi^4$ and $\phi^6$ Models,
	\newblock in {\em J. of Phys.: Conf. Series} {\bf 482}, 012045 (2014).
	
	\bibitem{weigel2019collective}
	H.~Weigel,
	\newblock Collective coordinate methods and their applicability to $\phi^4$
	models,
	\newblock in {\em A Dynamical Perspective on the $\phi^4$ Model}, p. 51--74,
	Springer, (2019).
	
	\bibitem{dorey2021res}
	P.~Dorey, A.~Gorina, I.~Perapechka, T.~Roma{\'n}czukiewicz, and Y.~Shnir,
	\newblock Resonance structures in kink-antikink collisions in a deformed
	sine-Gordon model,
	\newblock JHEP {\bf 2021}, 1--38 (2021).
	
	\bibitem{sutcliffe2022bound}
	P.~Sutcliffe,
	\newblock Boundary metrics on soliton moduli spaces,
	\newblock JHEP {\bf 2022}, 1--11 (2022).
	
	\bibitem{CenFring}
	J.~Cen and A.~Fring,
	\newblock Complex solitons with real energies,
	\newblock J. Phys. A: Math. Theor. {\bf 49}, 365202 (2016).
	
	\bibitem{CorreaFring}
	F.~Correa and A.~Fring,
	\newblock Regularized degenerate multi-solitons,
	\newblock JHEP {\bf 2016}, 8 (2016).
	
	\bibitem{fring2020BPS}
	A.~Fring and T.~Taira,
	\newblock Complex BPS solitons with real energies from duality,
	\newblock J. of Phys. A: Math. Theor. {\bf 53}, 455701 (2020).
	
	\bibitem{Urubu}
	F.~G. Scholtz, H.~B. Geyer, and F.~Hahne,
	\newblock Quasi-Hermitian Operators in Quantum Mechanics and the Variational
	Principle,
	\newblock Ann. Phys. {\bf 213}, 74--101 (1992).
	
	\bibitem{Bender:1998ke}
	C.~M. Bender and S.~Boettcher,
	\newblock Real Spectra in Non-Hermitian Hamiltonians Having PT Symmetry,
	\newblock Phys. Rev. Lett. {\bf 80}, 5243--5246 (1998).
	
	\bibitem{Alirev}
	A.~Mostafazadeh,
	\newblock Pseudo-Hermitian Representation of Quantum Mechanics,
	\newblock Int. J. Geom. Meth. Mod. Phys. {\bf 7}, 1191--1306 (2010).
	
	\bibitem{PTbook}
	C.~M. Bender, P.~E. Dorey, C.~Dunning, A.~Fring, D.~W. Hook, H.~F. Jones,
	S.~Kuzhel, G.~Levai, and R.~Tateo,
	\newblock PT Symmetry: In Quantum and Classical Physics,
	\newblock (World Scientific, Singapore)  (2019).
	
	\bibitem{BDodd}
	R.~Dodd and R.~Bullough,
	\newblock Polynomial conserved densities for the sine-Gordon equations,
	\newblock Proc. Royal Soc. of London. A. Math. and
	Phys. Sci. {\bf 352}(1671), 481--503 (1977).
	
	\bibitem{zhiber1979klein}
	A.~V. Zhiber and A.~B. Shabat,
	\newblock The Klein--Gordon equation with nontrivial group,
	\newblock in {\em Doklady Akademii Nauk} {\bf 247}, 1103--1107 (1979).
	
	\bibitem{correa2022linearly}
	F.~Correa, A.~Fring, and T.~Taira,
	\newblock Linearly stable and unstable complex soliton solutions with real
	energies in the Bullough-Dodd model,
	\newblock Nucl. Phys. B {\bf 979}, 115783 (2022).
	
	\bibitem{andreevback}
	V.~A. Andreev,
	\newblock B{\"a}cklund transformations of the Bullough-Dodd-Zhiber-Shabat
	equation and symmetries of integrable equations,
	\newblock Theor. and Math. Phys. {\bf 79}, 448--450 (1989).
	
	\bibitem{Fring:1992pj}
	A.~Fring, G.~Mussardo, and P.~Simonetti,
	\newblock Form-factors of the elementary field in the Bullough-Dodd model,
	\newblock Phys. Lett. B {\bf 307}, 83--90 (1993).
	
	\bibitem{assis2008bullough}
	P.~E.~G. Assis and L.~A. Ferreira,
	\newblock The Bullough--Dodd model coupled to matter fields,
	\newblock Nucl. Phys. B {\bf 800}, 409--449 (2008).
	
	\bibitem{CCFsineG}
	J.~Cen, F.~Correa, and A.~Fring,
	\newblock Degenerate multi-solitons in the sine-Gordon equation,
	\newblock J. Phys. A: Math. Theor. {\bf 50}, 435201 (2017).
	
	\bibitem{anco2011}
	S.~C. Anco, N.~T. Ngatat, and M.~Willoughby,
	\newblock Interaction properties of complex modified Korteweg--de Vries (mKdV)
	solitons,
	\newblock Physica D: Nonlin. Phen. {\bf 240}, 1378--1394 (2011).
	
	\bibitem{cen2019asymptotic}
	J.~Cen and A.~Fring,
	\newblock Asymptotic and scattering behaviour for degenerate multi-solitons in
	the Hirota equation,
	\newblock Physica D: Nonlin. Phen. {\bf 397}, 17--24 (2019).
	
	\bibitem{cen2017time}
	J.~Cen, F.~Correa, and A.~Fring,
	\newblock Time-delay and reality conditions for complex solitons,
	\newblock J. of Math. Phys. {\bf 58}, 032901 (2017).
	
	\bibitem{pereira2021some}
	C.~F. Pereira, G.~Luchini, T.~Tassis, and C.~P. Constantinidis,
	\newblock Some novel considerations about the collective coordinates
	approximation for the scattering of $\phi^4$ kinks,
	\newblock J. Phys. A: Math. Theor. {\bf 54},
	075701 (2021).
	
\end{thebibliography}

\end{document}